# Wireless enabled clothing

New modular technologies and overall supply chain impact

L-F Pau

CBS, Copenhagen, Denmark;
lpau@nypost.dk

Erasmus University, Rotterdam, Netherlands

*Abstract*: **The paper is devoted to the realization of wireless enabled clothing, employing recent new technologies in electronics, textile, and renewable power. This new wireless enabled clothing architecture is modular and distributed, allowing for customization in functionality and clothing designs. Are studied the implications for supply chains, distribution channels, and cost benefits. Modular wireless enabled clothing offers significant personalization opportunities at costs comparable with mobile terminals.**

## I. INTRODUCTION

The present times exhibit a significant stabilization of the mobile terminals and smartphones uptake rate, still fueled mostly by wireless infrastructure, applications and electronic component evolution. Commodization processes apply to all three, without any rethinking of the core user terminal concept realization. At the same time, from the psychological and sociological sides, mobile terminals are becoming deeply ingrained into personal behaviors even if appearing as a device separate from the human body. Has time come to « hide the device » and instead embed the communications functionality into the closest interface to the body which is clothing, while capitalizing on substantial progresses made in some other technological fields than electronics and software? This would also significantly free human tactile and visual attentions, as shown by early prototypes in professional [1] or defense niche segments [2]. It would also fundamentally largely decouple the wireless usages from electrical power capacity pitfalls. The wireless clothing would become the human's second skin.

Next, if the communications functionality would be embedded into human clothing, completely different manufacturing and distribution channels would emerge, which have long achieved major skills in commodization as well as personalization (retail stores, platforms of e-tailers, fashion designs). A key question is whether the achievable production costs of communications functionality embedded in clothing would be competitive with those of the current mobile terminals industry.

This paper addresses the technological and cost feasibility of communications functionality embedded in clothing, revisiting largely unexploited recent technological breakthroughs in some other domains then consumer electronics, especially architecture, textile technology and renewable power. It is about how to free your hands and attention, so that you and your body alone can be extended with wireless communications and media access.

This present research is not about wearable computing, smart watches and wearables [3-4], medical smart fabrics, intelligent textiles [5], or Internet-of-Things. The connectivity to the Internet or the cloud for most of these wearables is still through a smartphone via a Wi-Fi, Bluetooth or ZigBee link, a configuration not addressed here.

To simplify terminology, « Wireless clothing » is the short-cut designation in this paper for modular wireless multi-network access embedded in clothing, with possible extensions beyond communications.

This research has addressed the required novel technologies, architecture, design, manufacturing and economic feasibility of wireless communications functionality embedded in clothing, taking into account innovations in that industry and real data. Therefore, it should be of no surprise to the reader that the references are dominated by technological realizations by niche start-ups, and by large collaborative industrial projects.

## II. MODULAR WIRELESS CLOTHING

### 2.1. Paradigm shifts in the wireless access terminal concept

Table 1 summarizes some recent user- and usage-driven paradigm shifts. It shows that wireless clothing is one way forward, besides gadgets like smartwatches and other specialized electronic communication devices coupled to the body (helmets, glasses, etc...). While it has specific limitations (Section 3.6), the special appeal of wireless clothing is that it has less of the drawbacks linked to user interface size and dependence on use of human attentive vision.

TABLE 1: User and usage driven paradigms in wireless access terminals.

| 1998-2008 | Since approx. 2008 |
|---|---|
| Wireless handheld phone with growing functionality | Wireless multi-service platform (communications, Internet, video, music, games, virtual reality…) |
| Homogeneous world products | Market or user segment differentiation, not the least on cost and brand |
| Competition with fixed phone and broadband access nodes | Competition from a wide set of platforms, incl. WiFi and gaming, with personal networking |

| Accessibility, Media and Style | Fashion, Function and Fun, which are also the symbolic and message bearing aspects of clothing |
| Addition to fixed phone with social and ubiquity advantages | Main social and consumer interaction platforms with social, dependability, health monitoring and personalization advantages |

### 2.2. Relevant technological achievements

A number of technological achievements or research results have been achieved over the few recent years, which all have a high relevance for wireless clothing. They are briefly described below in combination, as it is a portfolio of such technologies, and of pre-existing electronic components, which constitute the technology platform needed to realize wireless clothing.

#### A. Microelectronics

-Passive components embedded into rigid and flexible substrates [6];
-Vacuum roll-to-roll process for flexible OLED / Amoled displays; flexible or curved OLED displays (already commercialized) (Samsung Galaxy Round, LG Electronics G Flex 2);
-Research projects on foldable graphene based displays;
-OLED high contrast displays with photo emitting molecules and low consumption;
-Flexible LED dot arrays for wearable color announcements or push buttons (already commercialized);
-Miniature GPS/ GNSS antennas and circuits (already commercialized);
-Miniature encapsulated MEMS sensors with sub-threshold CMOS circuits (already commercialized);
-RFID passive tags for system configuration;
-Bluetooth smart (1 cm2 all-in-one for sensor, radio, antenna) (standardized and commercialized);
-Flexible 16/32 bit asynchronous processors via thin film transistors and nanowires ([7] and Epson);
-New know-how about on-body antennas and propagation;
-Folded ultra-thin camera lenses [8];
-Self-cleaning screen displays with nanoparticles;
-Super elastic electronic circuits using an AuGa film [9-10];
-Flexible Li-Ion battery (Samsung);
-Flexible flash memory with organic semiconductors (University of Tokyo).

#### B. Textiles

-Body heat thermogenerating textile fibers generating electrical power [11];
-Thermo-regulating fibers with polyactide adjusting to body temperature, or electronic sub-system temperature [13];
-Surface treatments to improve the hydrophobicity (water repelling), self -cleaning ability, and antimicrobial properties of 3D knitted insulating fabrics [13];
-Nanocrystals allowing for solar energy conversion while vaporized on a textile fiber or as a film;
-Touch sensitive textile fibers (already commercialized);
-Woven textile reinforced thermoplastics;
-eWall EMI/EMC protective textile against EM fields;
-Sticking markers and glues onto textile (already commercialized);
-Textiles with shape memory;
-Lighting OLED textile fibers (already commercialized) [14-15]; they either integrate LED's with conductive yarns, or have yarns coated with thermochromic ink;
-Flagrance diffusion from micro encapsulated textiles [16] and cosmetological textiles (Onixxa);
-Specialty chemicals for molds and protection of electronics modules instead of metal [17];
-3D printing of some categories of personalized clothing, using new fibers;
-Conductive cotton and semiconducting cotton fibers [18];
-Virtual simulation encompassing physical and mechanical properties of textiles for faster setup of manufacturing machines [19];
-Knitting technologies enabling multi-layer structures to be produced so that the textile is electrically insulated, except where it needs to be conductive [20];
-Textiles with silver particles offering EMI protection [21].

#### C. Renewable energy

-Non imaging solar flux concentrators;
-Photovoltaic embedded textile panels (already commercialized) [22-23];
-Photovoltaic textile fibers;
-Foldable solar panels on rolls [24];
-Small fuel microcells with H2O cartridges (already commercialized for smartphones).

### 2.3. Supply and distribution chain drivers

As they come from two different process, machinery and skills bases, it is necessary to contrast the supply chain drivers for current wireless terminal products, and the clothing industry, irrespective of technologies.

The clothing industry achieves volumes of over 60 Billion products worldwide per year. It is linked to the textile processing, textile manufacturing equipment, and chemical industries, all very decentralized. Especially the clothing and chemical industries have large value-added mass customization experiences. The distribution channels are either very distributed, or highly concentrated for some product lines, but all exhibit extremely high inventory rotation speeds. In general the customer led clothing renewal rates are indeed high. The clothing industry has very many customization channels, including clothing re-design, matching of accessories, and personalization / fitting [25]. The industry has very strong design and fashion differentiators, as well as very high value brands or no-brand approaches. Surprisingly, most of the clothing manufacturing (as opposed to fiber and yarn production) is by companies or communities located geographically in the same areas as electronics outsourcing manufacturing suppliers.

The wireless terminals manufacturing industry has volumes of about 2.6 Billion units/year (not including gaming terminals, WiFi devices, etc.). The producers are all linked to specific players in the electronics, IT, communications or Internet fields, and their channels. Large parts of the electronics assembly, some component and interconnect designs and test suites are outsourced (HTC, Foxtronn, Sanmina, Inventec, Wistron, Compal, Celestica, Flextronics, etc.). Customization, service creation, and content aggregation are largely concentrated. The distribution channels for terminals have rather low inventory rotation speeds, except at product launch times. The value-add from the terminals is often less than the value add from content, services or software applications. The user renewal rate of terminals is in general slowing down and in the 18-30 months intervals in different markets. About 70 % of products are from high value brands, often seeking diversifications, and mergers/acquisitions are frequent.

The textiles and clothing sector is a diverse and heterogeneous industry which covers a wide variety of products from hi-tech synthetic yarns to wool fabrics, cotton bed linen, to industrial filters or specialized industrial uniforms, or nappies to high fashion. This diversity of end products corresponds to a multitude of industrial processes, enterprises or market structures. Textiles and clothing account for around 4% of total manufacturing value added and 7% of manufacturing employment in the European Union. According to Euratex & Eurostat, the textile and clothing industry in the European Union had in 2012 a turnover of 186 BEuros, carried out investments of 5.9 BEuros, employed 2 M workers, and included 108 435 companies with an average 19 employees/company. The imports from outside European Union represented 70.2 % of turnover and exports 40.7 %, for a net deficit trade balance of (-29) BEuros.

### III. WIRELESS CLOTHING PRODUCT CONCEPT , ARCHITECTURE AND RESULTING SUPPLY CHANGES

*3.1. Researched Product family concept*

The user buys in the same outlet a clothing item with its enabled embedded wireless multi-network access (except for IMEI or other identification). The user equipment architecture is assumed decentralized. Most electronic sub-systems for communications, storage and sensing, are modular, removable, replaceable, and customizable by software. Many designs exist for 3G/4G analog front ends and processors combined weighing less than 75 g. In turn, sub-systems exploiting textile properties, such as antenna [26], energy harvesting, buttons, light emitting diodes, and some wires, are embedded into the clothing, and these sub-systems are washable & ironable. Those sub-systems which are rigid in at least one direction, can be snapped or fixed internally or externally to the clothing at body locations which normally stay linear and do not affect the « looks » of the user: e.g. above shoulder bones, on lateral sides of lungs, at root of the body back at belt level, in a belt, and/or on forearm.

Personalization / adjustments are possible at two levels. Sub-systems linked to textile properties can serve creating motives or shapes on the clothing, driving branding and fashion trends; they may in addition carry body sensors. Removable sub-systems, mostly electronic, being modular, can be changed as long as the interconnects, drivers and the distributed operating systems so allow.

By extension, the above product family concept can be mirrored in similar functionality located in parts in shoes / socks [27]; shoes offer excellent piezoelectric energy harvesting capabilities.

*3.2. User interfaces*

The wireless clothing user interacts for communications purposes with the usual, but here customizable, set of interfaces: buttons touch sensitive display, head up display, microphone, earphone or equivalent, as well as other application specific interfaces:
-the buttons are embedded into the wireless clothing, and can be placed at many locations: wrist, chest, gloves, belt, depending on the nature or usage of the clothing and accessibility requirements;
-the touch sensitive display is an electronic sub-system, which can be scratched or snapped onto clothing (typically on forearm), or be a separately powered large screen display connected by Bluetooth smart to the graphic processor (like for Microsoft Lumia 950, or Airwave ski helmet [28]), or be a head-worn display [29];
-the listening function can be either via a small set of speakers attached to the clothing, or a headset (wired/wireless) , or wireless earphone plugs [30], or a headset for listening by bone conduction for high surrounding sound environments [31-32];
-the microphone can be placed at many clothing locations, or exploit bone vibrations [2].
Notifications can be switched between LED diode light emission, e.g. on wrist, or a vibration in the neck [33].

*3.3. Power*

Two complementary approaches are considered. In the autonomous mode, the clothing's electronics would be reloaded by photovoltaic energy capture via the clothing's' own fibers. As prototypes have shown, 10 m of photovoltaic fibers, or a 20 x 20 cm flexible photovoltaic panel, cover the consumption of a typical smartphone with ¼ of time in emitting mode (see e.g. the Ikini iPhone bikini loader [34]) A future alternative are piezoelectric ZnO nanomaterials energy-harvesting devices, exploiting the electromagnetic environment [35]; such fibers placed at elbows or knees would generate some power. Exact power analysis must be performed for each wireless clothing configuration and intended usage (esp. usage of the display).

In the auxiliary power mode, simple coat hangers can reload a piece of wireless clothing via a standard power interface.

*3.4. Distributed operating system*

Wireless clothing requires a distributed low power consumption multitasking operating system with low latency, especially if sensors are connected. In many ways, Linux, Android and TRON fulfill most requirements. But additional operating system modules may be necessary, to manage sub-system connectivity & synchronization, and interconnectivity with other sensors, displays and devices.

The Android and Android Wear SDK may cause inter-process communication and latency problems, which may be alleviated by context awareness middleware enhancements, and by dynamic adjustable buffering for better system level inter-process communication.

*3.5. Manufacturing costs, distribution and retail*

The branding of the wireless clothing product family is primarily linked to the distribution network associated with the sales outlet, to fashion design names, and to innovation features [36].

Mobile terminals distribution and retail chains are primarily vertical (terminals supplier owned/franchised, operator owned/franchised, electronic distribution chains); the cascade of margins in vertical chain relationships drive up prices, and diffuse customer focus.

On the contrary, wireless clothing distribution and retail operate for most sub-systems as horizontal diversified value-adding distribution networks. The sales and customization are by specialist or strongly branded clothing or fashion outlets. The choice given to users at final retail outlet stage for most modular sub-systems, allow to reduce the accumulation of margins in the manufactured product. In fact, this outlet takes most of its margin by packaging, integration and personalization services to make the specific piece of wireless clothing fit the needs, looks and affordability of each customer [16]. Another major advantage of clothing outlets is the high turn around in trendy low-cost fast-renewed and customized clothes (Inditex/Zara, H&M, Farfetch, etc.).

To illustrate the costs for a 3G+ 16 GB integrated mobile user equipment, the study [37] provides the following split of the manufacturing cost of 277,70 $ , corresponding to a typical sales price in the 550 $ range : display (65 $), glass cover and touch sensors (30 $), battery (21 $), CPU (19.50 $), SDRAM (7.30 $), NAND Flash (29.50 $), WiFi / Bluetooth (8.05 $), touchscreen control (5.50 $), audio codec (1.20 $), energy management (3.35 $), 3G+ baseband chip(s) (10 $), GPS (2.60 $), integration & testing (74.70 $).

To illustrate in comparison the costs for a normal piece of clothing, such as a regular pair of jeans, the study [38] provides the following split of the sales price of 95 Euros: 31.2 % for the rental and maintenance of the sales outlet, 8 % commissions on banking card payments, 27 % taxes, 6.5 % profit / royalty to the brand, 12.9 % for publicity & marketing, 0.5 % for transport & insurance, 4.6 % for tissue and other materials, 4.2 % for washing and packaging, 0.9 % for manufacturing labor force, 4.2 % for tissue cutting and weaving. The manufacturing cost is 13.8 Euros.

*3.6. Usage limitations and risks of wireless clothing*

The envisaged product family has its reliability equal or higher than compact electronic mobile terminals; especially the distributed architecture, wider antenna and eventual energy harvesting surfaces, provide performance and maintainability gains. Mechanical fracture risks are equal or lower, as long as the user doesn't fall or be exposed to shocks. Electromagnetic radiation exposure risks may be slightly higher, although the bigger antennas reduce the radiation concentration but at the risk of covering larger areas [39]; methodologies exist to assess such health risks for wearables [40], possibly leading to standards.

Body movements will in general not render possible the inclusion of imaging devices into wireless clothing, leaving this functionality to smartphones or digital cameras; progress in attentive vision and eye movement analysis may however allow to revisit this issue.

Different are the risks linked to lifetime issues: brittleness, sub-module replacement convenience, user configuration errors, and cleaning. Normally all textile rooted subs-systems should be washable but it remains to be seen if advanced textiles meet all daily life washability and ironing criteria (corrosion from soap, water temperature) [13]. A possible risk-reduction measure is to tag each washable wireless clothing item with a RFID tag which communicates a required washing profile with the washing machine.

Risks exist also at customization level, as clothing retail shops may not have needed high tech competence, although they may have high competence in fixing accessories, sewing, and coloring.

*3.7. Wireless clothing manufacturing and assembly*

For mobile terminals, the assembly is mostly carried out by electronics contract manufacturers, which sometimes become unbranded terminals suppliers by learning. They handle only large product series volumes, and carry out testing & inspection far from customers. Reverse logistics on the electronics in handsets is rare.

In contrast, for wireless clothing, the textile / clothing related sub-systems and clothing assembly are carried out by diversified low-tech clothing workshops, with medium to small product series volumes [41]. Tools exist whereby custom requirements about fibers and textile materials can be input up-front resulting in faster weaving machine setup [19]. The assembly of electronic sub-systems is much simplified, except for interconnects which however can be handled by proven insertion machines. Reverse logistics flows would be active for the specialized electronic sub-systems, increasing reuse. Local repairs and enhancement for textile parts would be done in distribution outlets, or in association with them, directly with the end customers.

*3.8. Supply chain shifts*

Based on the differences highlighted above, a number of supply chain shifts can be conjectured in relation to wireless clothing emergence. A survey of 110 product designers and distribution companies across North America, Asia and Europe, has been carried out to collect views on the most likely supply chain shifts. The highest probability scenario resulting from this survey is the following:

1) First adopters, would be clothing distribution networks and fashion brands which already offer high value customization services (e.g. Levi Strauss assisted by Modelabs , professional specialist clothing suppliers, etc...) soon to be followed by those offering extended ranges of accessories (VF Corp, H&M, Inditex);
2) Normal high customer retention marketing and publicity campaigns of top-branded clothing industry would get higher attention than most mobile terminals suppliers' campaigns because of the sheer diversity and renewal rate of clothing offerings;
3) Electronic subsystem suppliers would welcome high rotation supply contract suppliers to clothing assembly and distribution companies, allowing to bypass integrated mobile terminals vendors;
4) It would be a welcomed challenge given clothing assembly and textile manufacturers to move into higher valued products;
5) A wireless clothing tailor role, with assistance of special IT tools, is to be taken over by industrial design and fashion specialists, rivaling electronics distribution sales personnel.

In line with technology management theories, vertically integrated industries get exposed to fragmentation driven by end users, outsourcing and product/channel concepts from other industries (e.g.: car , car parts and automotive electronics). They also get exposed to internal inefficiencies and slower platform evolution. It is seen that low-to-middle end mobile terminals vertical companies are already now exposed to fragmentation; this has happened because of platform and systems complexities, and of the growing power of independent product integrators sourcing subsystems from multiple competing sources. High end smartphones produced by vertically controlled supply chains or single companies, merge in the face of very high design and integration platform complexities and costs.

On the contrary, low cost, high flexibility agile product integrators with powerful pre-existing channels (such as textile, clothing, and specific functionality devices / ASIC's), betting on standardized technologies, have specific opportunities rooted in a stronger interaction with the end customers [19].

Two distinct business models need therefore to be addressed for mobile clothing, independently from the corresponding common business process described in Section 4. One business model is the extension of an existing customization model of an international clothing brand. The other is a mass customization model targeted at SME's, or micro-factories. The benefits of mass customization are that it combines the personalization of custom made products with the flexibility and cost efficiency of mass production [25].

## IV. INTEGRATION PROCESS MODELING FOR WIRELESS CLOTHING

### 4.1. Integration process modeling

Taking a comparative approach, this research has designed formal integration business process models (BPM) with parametric estimates or assumptions for both mobile terminals as well as for wireless clothing. They are rooted in the supply chain steps described in the previous Sections, and in bill-of-materials or -process cost data provided by industrial partners and Fashion To Future project [42]. Assuming in both cases the same 3G+ based LTE functionality, were identified: the integration process workflows for mobile terminal user equipment's, the integration process workflow for wireless clothing, as well as the specialized textile and clothing steps. It was shown that for mobile terminals, the operators and application providers are the sole personalization agents; for wireless clothing, mass customization comes mostly from the differentiated textile subsystems, the electronic subsystems being personalized almost as for mobile terminals. As a result, wireless clothing offers a diversity of physical, esthetical and functional service bundles.

### 4.2. Manufacturing cost and margin implications

In a second stage, the previous BPM models allow to carry out full impact analysis for end customers, industrial players, throughputs, net manufacturing margins, but not on upstream R&D. For vertically integrated mobile terminals distribution, the end-distributors expect the same high margins as the initial industrial product designer which is in a monopolistic position. For modular wireless clothing, clothing distributors compete both on visual designs, as well as on sub-systems selection/ suppliers; in this way the end distributors can impose their margins on all distributed and competing subsystem suppliers. Table 2 shows the different mixes in design costs, manufacturing costs, and distribution costs, while Table 3 shows the product price and profit implications in Euros/ product unit (excluding taxes), for the same 3G+ LTE functionality and performances. It is shown in this example not only that the sales price of the mobile terminal is higher, but also that the wireless clothing distribution value chains has vastly higher service revenues.

## V. CONCLUSION

The researched wireless enabled clothing concept forces designers and suppliers to think again over the whole communications access functionality and how it fits into people lives. It hinges on the distribution of functionality amongst a set of cooperating specialties, which together provide more value and user friendliness than the sum of the parts. Wireless clothing is a powerful re-intermediation agent of current mobile platform supply chains. It offers strong branding and customization possibilities at low cost to end users.

Textile and clothing industries , starting with specialist clothing and fashion, may become small but significant change agents while reaping some of the rewards mobile terminal suppliers may lose .This trend can only be slowed down by less capital and innovation funding in textile and clothing industries

This paper did not address the new capabilities offered by wireless clothing, e.g. gestures may be sensed by the clothing we wear and used to control some applications, and new biometric identification schemes may be deployed. This new skin will protect the body from the outside, but also heal and enhance it as clothing has also symbolic and message bearing aspects.

ACKNOWLEDGMENT

This research was sponsored by COST 605, COST 1302, Consortium "Fashion for the future", and three venture funds.

TABLE 2: MIXES IN DESIGN, MANUFACTURING AND DISTRIBUTION COSTS

**Mobile terminal**

| SUPPLIERS | Share in design | Share in manufacturing | Share in distribution |
|---|---|---|---|
| Mobile terminal platform & testing | 35 | 22 | |
| RT Software | 15 | 12 | |
| Applications software | 10 | 6 | |
| Displays | 8 | 12 | |
| Microelectronics circuits & assembly | 24 | 49 | |
| Mechanics & power | 6 | 16 | |
| Packaging | 2 | 5 | |
| | 100 | 100 | |
| CHANNELS | | | |
| Operators | | | 60 |
| Electronics shops | | | 30 |
| Enterprise applications | | | 10 |

**Wireless clothing**

| SUPPLIERS | Share in design | Share in manufacturing | Share in distribution |
|---|---|---|---|
| Textile materials | 3 | 6 | |
| Clothing manufacturing | 8 | 20 | |
| Mobile terminal platform | 20 | 15 | |
| RT Software | 17 | 14 | |
| Applications software | 8 | 5 | |
| Displays | 10 | 14 | |
| Microelectronics circuits & assembly | 20 | 10 | |
| Mechanics & power | 11 | 13 | |
| Pressing & Packaging | 3 | 3 | |
| | 100 | 100 | |
| CHANNELS | | | |
| Agents | | | 40 |
| Shops, eShops, Franchises | | | 45 |
| Enterprise dresses & applications | | | 15 |

TABLE 3: PRODUCT PRICE AND FUNCTIONALITY

| Euros | Mobile terminal | Wireless clothing |
|---|---|---|
| Electronics Manufacturing | 80 | 65 |
| El. Manufacturing profit | 40 | 19,5 |
| % | 50 | 30 |
| Textile and clothing | | 15 |
| Profit on clothing | | 7,5 |
| % | | 50 |
| Distribution cost | 55 | 35 |
| Distribution profit | 16,5 | 10,5 |
| % | 30 | 30 |
| Inventory rotation days | 45 | 15 |
| Enhancements | | 40 |
| Accessories | 30 | |
| **SALES PRICE** | **221,5** | **192,5** |
| Cost of goods sold | 105 | 120 |
| **PROFITS Euros** | Electronics manufacturer | Clothing manufacturer |
| | 35,6 | 13,16 |
| | Clothing manufacturer | Electronics manufacturer |
| | 0 | 13,85 |
| | Distribution Channel | Distribution Channel |
| | 16,5 | 10,5 |
| | Service revenue | Service revenue |
| | 0 | 40 |

REFERENCES

[1] OSCAR communicating helmet detecting dangerous AC currents at a distance, COLAS Rail, http://www.csgriboldi.com/oscar-un-nouveau-casque-de-chantier/
[2] « Félin » integrated combatant system, SAFRAN, http://www.safran-group.com/fr/defense/modernisation-du-combattant
[3] J. Wei, How wearables intersect with the cloud and the Internet of things, IEEE Consumer electronic magazine, July 2014, 53-56
[4] Wearable Technologies AG online, www.wearable-technologies.com
[5] Smart fabrics, Science Daily, Nov. 4, 2008, https://www.sciencedaily.com/releases/2008/10/081018191929.htm
[6] R. Carpenter, J. Hickman, Embedded passives: an emerging technology, Passive Component Industry., September 2004, 24-25
[7] D. Holman, R. Vertegaal, Organic user interfaces: designing computers in any way, shape, or form, Communications of the ACM, 51(6), 48-55
[8] A.L. Narayan, Folded optics enable ultrathin camera lens, OLE, March 2007, 11, www.optics.org/ole
[9] H. O. Michaud, J. Teixidor, S. P. Lacour, " Soft flexion sensors integrating stretchable metal conductors on a silicone substrate for smart glove applications", 28th IEEE International Conference on Micro Electro Mechanical Systems (MEMS), Estoril, Portugal, 2015
[10] Project PLACE-IT (Platforms for large area conformable electronics by integration), http://cordis.europa.eu/project/rcn/93792_en.html, and www.ohmatex.dk
[11] Eolane starts smart textiles (in French), Les Echos Entreprise, 2 October 2012, 20, http://www.lesechos.fr/02/10/2012/LesEchos/21282-097-ECH_eolane-se-lance-dans-le---smart-textile--.htm
[12] Project NOTEREFIGA, http://extra.ivf.se/noterefiga/template.asp
[13] Project SAFEPROTEX (High protective clothing for complex emergency operations), http://cordis.europa.eu/result/rcn/143604_en.html
[14] K. Cherenack, K. van Os, L. van Pieterson, Smart photonic textiles begin to weave their magic, Laser Focus World, April 2012, 63-66, www.laserfocusworld.com
[15] Project Texoled on OLED textile fibers, http://www.izm.fraunhofer.de/content/dam/izm/en/documents/Abteilungen/System_Integration_Interconnection_Technologies/TexLab/TexOLED.pdf
[16] J. Brinton, The fabulous dressing room (in French), Courrier International, no 1311, 17 December 2016, 48-49
[17] EMS Chemie, http://www.ems-group.com/de/produkte-maerkte/maerkte/elektro-elektronik/
[18] N. Savage, Electronic cotton, IEEE Spectrum, January 2012, 13
[19] Project MODSIMTEX, http://www.modsimtex.eu
[20] S. Harris, Catwalk goes techno, Engineering and Technology, 3(18), 25 October 2008, 27-29, www.theeit.org/engtechmag
[21] eWall EMI protection textile, www.ewall.fr
[22] Wypsips Crystal, SunPartner, http://www.objetconnecte.com/laibao-sunpartner-2905/
[23] Project DEPHOTEX, http://www.dephotex.com
[24] Solar cells on rolls (in Norwegian), SINTEF Materialer og kjemi, Gemini, no 2, December 2014, 8-10
[25] Project MICRO-DRESS (Customized wearable functionality and eco-materials extending the limits of apparel), http://cordis.europa.eu/result/rcn/140764_en.html
[26] P. S. Hall, Y. Hao, Antennas and propagation for body-centered wireless communications, ARTECH House, 2006, ISBN: 978-1-58053-493-2
[27] Sensoria, Communicating socks, http://www.sensoriafitness.com
[28] Oakley, Airwave connected ski mask, www.oakley.com/airwave
[29] G. Overton, Head-worn displays: useful tool or niche novelty, Laser Focus World, July 2013, 33-37 www.laserfocusworld.com
[30] Wireless smart Bluetooth earplugs by e.g. PK Paris, Motorola, Samsung, Apple
[31] ELNO, Headset for bone conduction, www.elno.fr/en
[32] Sonitus, Medical tooth bone based acoustic transmission, US Patent 7269266 B2, http://www.google.com/patents/US7269266
[33] Vest operates as a crisis mobile handset (in Norwegian), SINTEF Gemini, no 1, June 2014, 45, and EU Project "Societies"
[34] iKini bikini loader, http://www.gentside.com/bikini/ce-bikini-en-panneaux-solaires-recharge-votre-mp3_art25930.html
[35] Project PING (Piezoelectric nanogenerators on suspended microstructures for energy harvesting), http://cordis.europa.eu/result/rcn/142987_en.html
[36] Haute tech couture, Engineering and Technology, 3(18), 25 October 2008, 20-26
[37] iSupply, Estimate of 3G mobile phone material costs, IHS / iSuppli Corp., reproduced in Finanz & Wirtschaft, 29 May 2010, no 41, 35, https://technology.ihs.com/
[38] Pro-Tejer, Argentina, reproduced in Courrier International, no 1301, 8 October 2015, 41, http://www.fundacionprotejer.com/
[39] Z.N. Chen, A. Cai, T.S.P. See, X. Qing, M.Y.W. Chia, Small planar UWB antennas in proximity of the human head, IEEE Trans. Microwave theory and techniques, 54(4), 1846-1857
[40] S. Kirk, The wearables revolution is standardization a help or a hindrance, IEEE Consumer electronics magazine, October 2014, 45-50
[41] T-M Choi, Fashion supply chain management: industry and business analysis, IGI Global, 2012, ISBN: 978-1-60960-756-2
[42] Project Fashion to future, http://cordis.europa.eu/project/rcn/83475_en.html